\documentclass[11pt,oneside,letterpaper]{article}
\usepackage{amssymb}
\usepackage{amsmath}
\usepackage[dvips]{graphicx}
\usepackage{setspace}
\usepackage{fancyhdr}
\usepackage{ifpdf}
\usepackage{graphicx}
\usepackage{comment}

\def\half{{\textstyle{1\over2}}}

 \def\p{\partial}

\newcommand{\bea}{\begin{eqnarray}}
\newcommand{\eea}{\end{eqnarray}}
\newcommand{\be}{\begin{equation}}
\newcommand{\ee}{\end{equation}}

\newcommand{\bi}{\begin{itemize}}
\newcommand{\ei}{\end{itemize}}

\numberwithin{equation}{section}
\allowdisplaybreaks  

\begin{document}

\vspace*{2.5cm}
\begin{center}
{ \LARGE \textsc{{{Holography at an Extremal De Sitter Horizon}}}\\}
\begin{center}
Dionysios Anninos and Thomas Hartman
\end{center}
\end{center}
\vspace*{0.6cm}
\begin{center}
Jefferson Physical Laboratory, Harvard University, Cambridge, MA 02138, USA
\vspace*{0.8cm}
\end{center}
\vspace*{1.5cm}
\begin{abstract}
\noindent Rotating maximal black holes in four-dimensional de Sitter space, for which the outer event horizon coincides with the cosmological horizon, have an infinite near-horizon region described by the rotating Nariai metric.  We show that the asymptotic symmetry group at the spacelike future boundary of the near-horizon region contains a Virasoro algebra with a real, positive central charge. This is evidence that quantum gravity in a rotating Nariai background is dual to a two-dimensional Euclidean conformal field theory. These results are related to the Kerr/CFT correspondence for extremal black holes, but have two key differences: one of the black hole event horizons has been traded for the cosmological horizon, and the near-horizon geometry is a fiber over dS$_2$ rather than AdS$_2$.
\end{abstract}

\newpage
\setcounter{page}{1}
\pagenumbering{arabic}

\section{Introduction}

The cosmological horizon of de Sitter space resembles a black hole horizon in many ways, exhibiting a thermal spectrum and having entropy proportional to the horizon area \cite{Gibbons:1977mu}. On the other hand, it is sharply different from a black hole horizon in the sense that it is observer dependent, and its size does not decrease upon emission of Hawking radiation as it reabsorbs its own radiation. An important open problem is to understand whether there is a sensible theory of quantum gravity in a de Sitter background, and to account for its entropy from a microscopic point of view. There have been various attempts to do so in two and three dimensions \cite{Strominger:2001pn,Hawking:2000da,Banados:1998tb}, and in four dimensions \cite{Banks:2003cg,Banks:2005bm,Banks:2006rx}. The goal of this paper is to study the de Sitter horizon in the special case that it is coincident with a black hole horizon.

In what follows, we consider rotating black holes in asymptotically de Sitter spacetime in four dimensional Einstein gravity. A rotating black hole in de Sitter space has an inner, outer and cosmological horizon. Extremal Kerr-de Sitter black holes are obtained by bringing together the inner and outer horizons. The near horizon geometry of this configuration is given by a generalization of the near-horizon extreme Kerr (NHEK) spacetime \cite{Bardeen:1999px}, which is a fibered product of two-dimensional anti-de Sitter space and the two-sphere with an $SL(2,\mathbb{R})\times U(1)$ isometry group.

It was discovered in \cite{Guica:2008mu} that there exist consistent boundary conditions on NHEK for which the asymptotic symmetry group \cite{Brown:1986nw} enhances the $U(1)$ isometry to a Virasoro algebra with non-trivial central charge. This led to the conjecture that quantum gravity in the near horizon geometry is holographically dual to a 2d CFT.  In support of this conjecture, the central charge of the Virasoro algebra and the Frolov-Thorne temperature \cite{Frolov:1989jh} were used in the Cardy formula to account for the black entropy from the viewpoint of the proposed dual CFT.  This was extended to non-zero cosmological constant in \cite{Lu:2008jk,Hartman:2008pb}. For both positive and negative cosmological constant, the near horizon geometry is a fiber over AdS$_2$.

There is another extremal limit for rotating black holes in de Sitter space: the limit where the outer and cosmological horizons coincide. In this limit, the near horizon geometry becomes a fibered product of \textit{two-dimensional de Sitter space} and a two-sphere, with the metric
\be\label{intrometric}
ds^2 = \Gamma(\theta)\left[-(1-r^2)d\tau^2 + {dr^2\over 1-r^2} + \alpha(\theta)d\theta^2\right] + \gamma(\theta)(d\phi + krd\tau)^2 \ .
\ee
The functions $\Gamma,\gamma,\alpha$ and $k$ are given below in terms of the black hole parameters. This spacetime, which was extensively studied in \cite{Booth:1998gf}, is known as the rotating Nariai geometry and reduces to the Nariai geometry $dS_2 \times S^2$ \cite{nariai,Ginsparg:1982rs,Cardoso:2004uz} in the non-rotating case. Observers in the rotating Nariai spacetime live inside a cosmological horizon whose entropy corresponds to the entropy of the original horizon in the full geometry.

We will see that the rules developed for the NHEK geometry are also applicable to the rotating Nariai geometry. The asymptotic symmetry algebra of (\ref{intrometric}) at future spacelike infinity is the Virasoro algebra with real, positive central charge
\be\label{introc}
c_L = {12 r_c^2 \sqrt{(1-3r_c^2/\ell^2)(1+r_c^2/\ell^2)}\over -1 + 6 r_c^2/\ell^2 + 3 r_c^4/\ell^4} \ ,
\ee
where $3/\ell^2$ is the cosmological constant and $r_c$ is the cosmological horizon. We therefore conjecture that quantum gravity in the rotating Nariai geometry is dual to a Euclidean 2d conformal field theory. Assuming the Cardy formula, this together with the left-moving temperature allow for a holographic derivation of the cosmological horizon entropy.\footnote{In this extremal (or maximal) limit, what we call the cosmological entropy is of course also the black hole entropy.  Nonetheless the cosmological nature of the double horizon leads to a dual CFT qualitatively different from that dual to ordinary extreme black holes.}  While the derivation of the asymptotic symmetry group and central charge is robust, this application of the Cardy formula is on a speculative footing.  Unlike in AdS/CFT or Kerr/CFT, an understanding of how to map thermal states between the bulk and boundary is lacking.

Despite the apparent similarities to Kerr/CFT, there are important differences between NHEK and rotating Nariai. Asymptotically, the rotating Nariai spacetime is naturally foliated by a timelike ``radial" coordinate and the foliations are spacelike. This is in contrast to the timelike foliations of the NHEK geometry, suggesting that the appropriate conformal field theory lives on a spacelike manifold and is in fact Euclidean very much in the same vein as the dS/CFT proposal. Furthermore, it is unclear whether quantum gravity in the rotating Nariai geometry is unitary and thus application of the Cardy formula remains on a numerological footing much like it was in \cite{Bousso:2001mw}, where the entropy of asymptotically $dS_3$ conical defects was examined. Therefore we stress that this does not provide a firm microstate counting of the de Sitter entropy. On the other hand, the fact that a naive application of Cardy's formula indeed accounts for the cosmological entropy is a clear invitation for a solid explanation.

The rotating Nariai/CFT duality proposed here shares some characteristics with Kerr/CFT and dS/CFT, but is not continuously connected to either of these, and should be considered a separate class of dualities.  Adjusting the black hole mass to bring the inner black hole horizon toward the coincident outer and cosmological horizons, it seems possible to interpolate between Kerr/CFT and Nariai/CFT.  However, in the ultracold limit where all three horizons coincide, the near horizon geometry has no apparent CFT interpretation.  As in dS/CFT, the dual to rotating Nariai is Euclidean. In contrast, it is always two-dimensional, regardless of the dimensionality of the bulk spacetime.

Various generalizations, applications, and tests of the rotating Nariai/CFT correspondence are possible.  As a first step, the derivation of the asymptotic symmetry group is generalized to include electromagnetic charge in appendix \ref{charged}.  The extension to higher-dimensional de Sitter space is not considered here but should be straightforward (see \cite{Lu:2008jk}).  Since the near-horizon geometry in any number of dimensions is a fiber over $dS_2$, we expect the CFT to always be two-dimensional.  Another open question is how the correspondence applies to near-extremal black holes; progress along these lines in Kerr/CFT has been made in \cite{Castro:2009jf,Bredberg:2009pv,Amsel:2009ev}.

A natural way to test the rotating Nariai/CFT correspondence and develop the holographic dictionary would be to examine instabilities, from both the gravity and CFT sides of the duality.  On the gravity side, it would be interesting to study the classical stability of the rotating Nariai solution along the lines of \cite{Amsel:2009ev,Dias:2009ex}. Understanding the quantum instabilities of the rotating Nariai geometry (see \cite{Belgiorno:2009pq} for example) would also be worthwhile.  Instabilities must have a dual description in the CFT.  A similar motivation has led to non-trivial tests of the Kerr/CFT correspondence, where the instability on the gravity side is related to black hole superradiance, which has a dual description in terms of CFT two-point correlators \cite{Bredberg:2009pv,Cvetic:2009jn,Hartman:2009nz}.  In rotating Nariai/CFT, on the other hand, the potential instabilities on the gravity side are qualitatively different, so the dual description could provide new insight into the properties of the CFT.

We should mention an interesting observation concerning Kerr/CFT and the dimensionality of the quantum de Sitter Hilbert space
\cite{Banks:2003cg,Banks:2005bm,Banks:2006rx,Bousso:2000nf,Bousso:2002fi,Witten:2001kn,Goheer:2002vf}, before delving into our analysis. One may initially suspect that an extremal rotating black hole in de Sitter space with coincident inner and outer horizons, which has a near horizon geometry dual to a full chiral CFT, must have an infinite dimensional Hilbert space. On the other hand, such black holes - even though in thermodynamic equilibrium - are \emph{not} thermodynamically stable due to the incoming thermal radiation of the cosmological horizon that causes them to heat up \cite{Bousso:1999ms,Bousso:1997wi}, thus eliminating the original suspicion.

The outline is as follows. In section \ref{geom} we describe the geometry and thermodynamics of Kerr-de Sitter space and the rotating Nariai geometry. In section \ref{rot} we study the asymptotic symmetries of the rotating Nariai geometry and propose the rotating Nariai/CFT correspondence.  In particular, we apply the Cardy formula to the CFT and find that it reproduces the cosmological horizon entropy. Finally, we briefly discuss scalars in the rotating Nariai geometry in section \ref{scalars}.  In appendix \ref{charged} we show that the rotating Nariai/CFT correspondence generalizes to maximal dyonic Kerr-Newman-dS black holes.

\section{Geometry and Thermodynamics}\label{geom}

Our story begins with a review of the Kerr-dS and rotating Nariai geometries as solutions of Einstein gravity with a positive cosmological constant $\Lambda = 3/\ell^2$ in four-dimensions:
\be
S_{grav} = \frac{1}{16\pi}\int_{\mathcal{M}} d^4 x \sqrt{-g} \left(R - 2\Lambda\right) \ ,
\ee
where we have set Newton's constant $G=1$.
The Kerr-dS metric is
\be
ds^2 = -\frac{\Delta_{\hat{r}}}{\rho^2}\left(  d\hat{t} - \frac{a}{\Xi} \sin^2\theta d\hat{\phi} \right)^2 + \frac{\rho^2}{\Delta_{\hat{r}}}d\hat{r}^2 + \frac{\rho^2}{\Delta_\theta}d\theta^2 + \frac{\Delta_\theta}{\rho^2}\sin^2\theta \left( a d\hat{t} - \frac{\hat{r}^2 + a^2}{\Xi} d\hat{\phi} \right)^2
\ee
where
\bea
\Delta_{\hat{r}} &=& (\hat{r}^2 + a^2)(1-\hat{r}^2/\ell^2)-2 M \hat{r}, \quad \Delta_\theta = 1+a^2\cos^2\theta/\ell^2\\
 \rho^2 &=& \hat{r}^2 + a^2 \cos^2\theta, \quad \Xi = 1+a^2/\ell^2 \ .\notag
\eea
The three horizons are given by the positive solutions of $\Delta_{\hat{r}} = 0$ and we denote them by $r_-$, $r_+$ and $r_c$, with $r_c \geq r_+ \geq r_-$. There are three extremal limits one can consider, all with zero Hawking temperature: Taking the inner and outer horizon to coincide (extremal limit) such that $r_+ = r_-$, taking the outer and cosmological horizon to coincide (Nariai limit) such that $r_+ = r_c$, and taking the inner, outer and cosmological horizons to coincide (ultracold limit) such that $r_+ = r_- = r_c$. Each of these extremal configurations has a different near horizon geometry.\footnote{It is amusing to note that these near horizon geometries at fixed polar angle are all present in topologically massive gravity \cite{Anninos:2008fx,Anninos:2009zi,Anninos:2009jt}, where they were called warped AdS, warped dS, and warped flat space respectively.}

The extremal limit $r_+ = r_-$ was studied in the context of Kerr/CFT in \cite{Lu:2008jk,Hartman:2008pb}.  Here we instead focus on the Nariai limit $r_+ = r_c$. The Penrose diagram of the geometry in this limit is given in appendix \ref{Penrose}. The parameters $a,M$ in the Nariai limit are
\bea
a^2 &=& {r_c^2(1-3r_c^2/\ell^2)\over 1 + r_c^2/\ell^2}\\
M &=& {r_c(1-r_c^2/\ell^2)^2\over 1 + r_c^2/\ell^2} \ . \notag
\eea

\subsection{Thermodynamics of the Cosmological Horizon}

The thermodynamic properties of the Kerr-de Sitter spacetime were obtained in \cite{Dehghani:2002np,Ghezelbash:2004af}. The conserved charges associated to the $\partial_{\hat{t}}$ and $\partial_{\hat{\phi}}$ Killing vectors are given by
\be
\mathcal{Q}_{\partial_{\hat{t}}} = - \frac{M}{\Xi^2}, \quad \mathcal{Q}_{\partial_{\hat{\phi}}} = -\frac{a M}{\Xi^2} \ .
\ee
Since slices of constant $r$ are asymptotically spacelike, the charges are evaluated as integrals over $\hat{\phi},\theta$ at future spacelike infinity, and are ``conserved" in the sense that they do not depend on $\hat{t}$. The entropy of the cosmological horizon is given by the usual Bekenstein-Hawking relation
\be
S_{cosm} = \frac{\mbox{Area}}{4} = \frac{\pi(r_c^2 + a^2 )}{\Xi}
\ee
where we have set $G=1$. The first law of thermodynamics for the cosmological horizon becomes
\be
d \mathcal{Q}_{\partial_{\hat{t}}} = T_H d S_{cosm} + \Omega_H d \mathcal{Q}_{\partial_{\hat{\phi}}}
\ee
where $T_H$ is the Hawking temperature of the cosmological horizon and $\Omega_H$ is the angular velocity at the horizon. In the Nariai limit $r_+ = r_c$ the variation of the entropy can be expressed completely in terms of a variation of the angular momentum and takes the following form:
\be
d S_{cosm} = \beta_L d \mathcal{Q}_{\partial_{\hat{\phi}}}
\ee
where $\beta_L = 1/T_L$ is the chemical potential associated to the angular momentum. We refer to $T_L$ as the left moving temperature. Explicitly one finds
\be
T_L = \frac{(r_c^2 + a^2)^2}{4\pi a r_c \Xi}\frac{(6r_c^2/\ell^2 +3 r_c^4/\ell^4-1)}{(r^2_c + a^2)(1+r_c^2/\ell^2)}\label{firstlaw} \ .
\ee

\subsection{Rotating Nariai Geometry}

The rotating Nariai geometry \cite{Booth:1998gf} is obtained when the cosmological and outer horizons coincide, $r_c = r_+$. In this case, the Kerr-de Sitter solution becomes time dependent both in the region outside the cosmological horizon as well as the region between the cosmological horizon and the inner horizon.

We will take the Nariai limit $r_+\to r_c$ and the near horizon limit simultaneously. This is the Nariai analog of the near-NHEK limit of extremal black holes considered in \cite{Bredberg:2009pv,Castro:2009jf}. Define the non-extremality parameter
\be
\tau = {r_c - r_+ \over  r_c}  \ .
\ee
For small $\tau$, the Hawking temperatures of the outer and cosmological horizons are both
\be
T_H \approx {b \tau \over 4\pi} \ ,
\ee
where
\be\label{defb}
b = {r_c(r_c-r_-)(3r_c + r_-)\over \ell^2 (a^2 + r_c^2)}
 = {r_c(-1 + a^2/\ell^2 + 6 r_c^2/\ell^2)\over  (r_c^2 + a^2)} \ .
\ee
The near-horizon coordinates are
\be
t = b \lambda \hat{t} \ , \quad r = {r_c - \hat{r}\over \lambda r_c} \ , \quad \phi = \hat{\phi} - \Omega_c \hat{t} \ ,
\ee
where
\be
\Omega_c = {\Xi a\over r_c^2 + a^2}
\ee
is the angular velocity of the cosmological horizon. Taking $\lambda \to 0, \tau\to 0$ with $\tau/\lambda, t, r, \phi$ held fixed, we find the rotating Nariai metric \cite{Booth:1998gf}
\be\label{thermalnariai}
ds^2 = \Gamma(\theta)\left(r(r-\tau)dt^2 - {dr^2\over r(r-\tau)} + \alpha(\theta)d\theta^2\right) + \gamma(\theta)(d\phi + k r dt)^2 \ ,
\ee
with $\phi \sim \phi + 2\pi$, $r \in (0,\tau)$, and
\be\label{nariaiparam}
\Gamma(\theta) = { \rho_c^2 r_c\over b (a^2 + r_c^2)} \ , \quad
\alpha(\theta) = {b (a^2 + r_c^2)\over r_c \Delta_\theta} \ , \quad
\gamma(\theta) = {\Delta_\theta(r_c^2 + a^2)^2\sin^2\theta\over\rho_c^2 \Xi^2} \ ,
\ee
\be\notag
k = - {2 a r_c^2\Xi\over b (a^2 + r_c^2)^2}\ , \quad
\rho_c^2 = r_c^2 + a^2 \cos^2\theta\ .
\ee
The near horizon geometry is therefore a fiber over $dS_2$.  The coordinate change
\be
r \to {\tau \over 2}(r + 1) \ , \quad t \to {2\over \tau} t \ , \quad \phi \to \phi-kt
\ee
puts the $dS_2$ base into the familiar static coordinates
\be\label{staticc}
ds^2 = \Gamma(\theta)\left(-(1-r^2)dt^2 + {dr^2\over 1-r^2} + \alpha(\theta)d\theta^2\right) + \gamma(\theta)(d\phi + krdt)^2\ ,
\ee
with $r \in (-1,1)$. The isometry group of this spacetime is $U(1) \times SL(2,R)$ generated by
\bea
K_0 &=& \p_\phi \\
\bar{K}_0 &=& \p_t \notag\\
\bar{K}_1 &=& {r \sinh t\over\sqrt{1-r^2}}\p_t  + \cosh t\sqrt{1-r^2}\p_r - {k \sinh t \over \sqrt{1-r^2}}\p_\phi \notag\\
\bar{K}_2 &=& {r \cosh t\over\sqrt{1-r^2}}\p_t  + \sinh t\sqrt{1-r^2}\p_r - {k \cosh t \over \sqrt{1-r^2}}\p_\phi \ . \notag
\eea

\begin{figure}[t]
\begin{center}
 \input{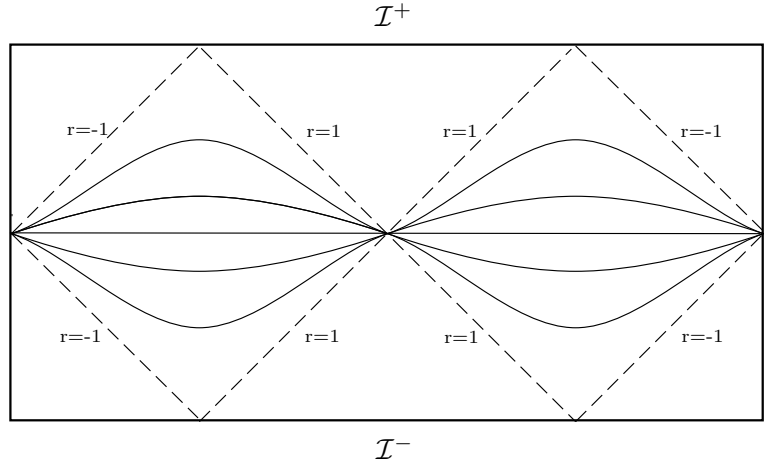}
 \end{center}
 \caption{\small Penrose diagram of $dS_2$ with the cosmological horizons explicit. The left and right sides are identified.}
 \label{dspenrose}
\end{figure}
In the Penrose diagram shown in Fig. \ref{dspenrose}, static coordinates cover the patch inside the dotted lines.  The same metric with $r>1$ covers a patch outside the cosmological horizon, including a segment of the boundary.\footnote{This is equivalent to taking $\tau < 0$, $r \in (0, \infty)$ in the metric (\ref{thermalnariai}), which is obtained as the near-horizon limit of Kerr-dS starting outside the cosmological horizon rather than between the two horizons. This is the $dS_2$ analog of the thermal ``near-NHEK" coordinates \cite{Bredberg:2009pv}.}  Observers in the rotating Nariai geometry find themselves enclosed by a cosmological horizon at $r = \pm 1$. The entropy associated to the cosmological horizon of the rotating Nariai geometry is
\be
S_{Nariai} = {\mbox{Area}\over 4} = \frac{\pi (r_c^2 + a^2)}{\Xi} \ ,
\ee
which is precisely the cosmological entropy in the full Kerr-de Sitter geometry. The boundaries reside at $\mathcal{I}^\pm$ and are spacelike for all values of $\theta$. Note that constant $r$ hyperslices are spacelike for large $r$ and thus $r$ is a timelike coordinate for large enough values.

The global coordinates defined by
\bea
\tan(\eta/2) &=& \tanh\left(\half \sinh^{-1}[\sqrt{1-r^2}\sinh t]\right)\\
\cos\psi &=& r(\cosh^2 t - r^2 \sinh^2 t)^{-1/2}\notag\\
\varphi &=& \phi + {k\over 2} \log\left(\sin(\eta + \psi)\over \sin(\eta - \psi)\right)\notag
\eea
with $\psi \sim \psi + 2\pi$, $\eta \in (-\pi/2, \pi/2)$ cover all of $dS_2$, with metric
\be
ds^2 = \Gamma(\theta)\left({-d\eta^2 + d\psi^2\over \cos^2\eta} + \alpha(\theta)d\theta^2\right) + \gamma(\theta)(d\varphi + k \tan \eta d\psi)^2 \ .
\ee
This coordinate change is useful to understand the $dS_2$ Penrose diagram, but below we will work in the coordinates (\ref{staticc}).

\subsection{Ultracold Solution}

For completeness we also include the near horizon geometry in the ultracold limit where the inner, outer and cosmological horizons coincide \cite{Booth:1998gf}. (We do not know of any way to identify the ultracold geometry with a CFT.) In this limit, the parameter $b \propto r_c - r_-$ in (\ref{defb}) and (\ref{nariaiparam}) goes to zero, so we must rescale coordinates appropriately.  Starting with the static metric (\ref{staticc}) on rotating Nariai, defining the ultracold coordinates
\be
r = \tilde{r}\sqrt{b} \ , \quad t = \tilde{t}\sqrt{b} \ ,
\ee
and taking $b \to 0$ with the ultracold coordinates held fixed, we find
\be
ds^2 = \tilde{\Gamma}[\theta]\left[ -d\tilde{t}^2 + d\tilde{r}^2 + \tilde{\alpha}(\theta) d\theta^2 \right] + \gamma(\theta) \left( d\phi + \tilde{k} \tilde{r} d\tilde{t} \right)^2
\ee
with
\be
\tilde{\Gamma} = b\Gamma \ , \quad \tilde{\alpha} = \alpha/b \ , \quad \tilde{k} = bk \ .
\ee
The ultracold geometry is a fibered product of two-dimensional Minkowski space and the two-sphere. We can also obtain the solution as a fibered product of two-dimensional Rindler space and the two-sphere via a coordinate transformation.  Which metric on Minkowski space naturally appears in the ultracold limit depends on how the limit is taken starting from the full Kerr-dS black hole.

\section{The Rotating Nariai/CFT Proposal}\label{rot}

The Kerr/CFT correspondence proposes that quantum gravity in the NHEK geometry is dual to a thermal state in a 2d conformal field theory. The argument consists of imposing a set of boundary conditions for all excitations of the geometry, finding the asymptotic symmetry group satisfying these boundary conditions and computing any additional central charges that may be present.

We will now show that a similar argument allows us to relate quantum gravity in the rotating Nariai geometry to a 2d CFT. One of the differences between the proposal we make and that of Kerr/CFT is that the putative dual CFT lives on a spacelike manifold given that constant $r$ slices are in fact asymptotically spacelike. We take this as an indication that the dual CFT of the cosmological horizon is Euclidean. In this sense, the proposed duality lies along the same vein as the dS/CFT correspondence. Furthermore, given that there are two disconnected boundaries we are confronted with the question of where the dual conformal field theory resides. We will not address this question which has yet to be fully understood in the Kerr/CFT case.

The asymptotic symmetry group (ASG) of a theory is the set of allowed, nontrivial symmetries.  A symmetry is allowed if it generates a transformation satisfying the boundary conditions, and nontrivial if it falls off at infinity slow enough for the associated conserved charge to be nonzero on some background.  The ASG depends on the boundary conditions and on the action, which defines the conserved charges.  Generally for a given action the window of consistent boundary conditions is small; if one chooses boundary conditions that are too loose, the conserved charges diverge, while if the boundary conditions are too tight, then all excitations are gauge-equivalent to the vacuum and the theory is trivial.

\subsection*{Boundary Conditions}

We choose our boundary conditions in direct analogy to the NHEK boundary conditions proposed in \cite{Guica:2008mu}.  In the basis $(t,\phi,\theta,r)$,
\begin{equation}\label{metricbcbc}
h_{\mu\nu} \sim \mathcal{O}
\left(
  \begin{array}{cccc}
    r^2 & 1 & 1/r & 1/r^2 \\
     & 1 & 1/r & 1/r \\
     & & 1/r & 1/r^2\\
     &  &  & 1/r^3 \\
  \end{array}
\right)
\end{equation}
always keeping in mind that $r$ is asymptotically a timelike coordinate. Note that not all components of the perturbation are subleading with respect to the boundary metric. Below, we also impose a supplemental boundary condition that eliminates excitations above extremality.

\subsection*{Asymptotic Symmetries}
The boundary conditions are preserved by the diffeomorphisms
\bea
\zeta_\epsilon &=& \epsilon(\phi) \partial_\phi - r \epsilon'(\phi) \partial_r\\
\bar{\zeta} &=& \partial_\tau
\eea
Expanding in a basis $\epsilon_n = -e^{-in\phi}$, the left-moving diffeomorphisms $\zeta_\epsilon$ give rise to a single copy of the Virasoro algebra
\be
i[\zeta_n , \zeta_m] = (n-m)\zeta_{n+m} \ ,
\ee
with zero mode equal to the $U(1)$ isometry, $\zeta_0 = -\p_\phi$.

\subsection*{Charges}

For each diffeomorphism there is an associated conserved charge which can be computed using the Barnich-Brandt-Compere formalism \cite{Barnich:2001jy,Barnich:2007bf,Compere:2007az}. In particular, one finds the conserved charges
\be
Q_\zeta(h,{g}) = \int_{\mathcal{I}^+} k_{\zeta}[h;{g}] 
\ee
where $k_\zeta$ is constructed from the Einstein equations,
\begin{multline}
k_\zeta [h;g] = \frac{1}{4}\epsilon_{\alpha\beta\mu\nu} [\zeta^\nu D^\mu h - \zeta^\nu D_\sigma h^{\mu\sigma} + \zeta_\sigma D^\nu h^{\mu\sigma} \\  + \frac{1}{2} h D^\nu \zeta^\mu - h^{\nu\sigma} D_\sigma \zeta^\mu + \frac{1}{2} h^{\sigma\nu}(D^\mu \zeta_\sigma + D_\sigma \zeta^\mu)] dx^\alpha \wedge dx^\beta
\end{multline}
The integral is taken over the boundary of a constant-$t$ slice at $r \to \infty$ . To ensure a chiral spectrum and finite charges we impose the supplemental boundary condition $Q_{\partial_\tau}(h,{g}) = 0$.

The Dirac bracket algebra which dictates the algebra of asymptotic symmetries is given by
\be
\{ Q_{\zeta_m}, Q_{{\zeta_n}} \} = -i(m-n)Q_{\zeta_{m+n}} + \frac{1}{8\pi} \int k_{\zeta_m}[\mathcal{L}_{{\zeta_n}}\bar{g};\bar{g}]
\ee
Upon quantization, we transform the Dirac bracket algebra into a commutation relation allowing us to interpret the classical central charge as a quantum central charge of the dual conformal field theory. Replacing the classical charges $Q_{\zeta_m}$ by their quantum counterparts $L_m$, we find the Virasoro algebra,
\be
[L_m,L_n] = (m-n)L_{m+n} + \frac{c_L}{12}(m^3-m)\delta_{m+n}
\ee
with central charge
\be\label{cent}
c_L = 3|k|\int_0^\pi d\theta \sqrt{\Gamma(\theta)\alpha(\theta)\gamma(\theta)} \ .
\ee
Explicitly, the central charge is
\be
c_L = {12 a r_c^2 \over b (a^2 + r_c^2)} \ ,
\ee
which equals (\ref{introc}).  Note that $c_L$ is real and positive.

We therefore propose that quantum gravity in the rotating Nariai geometry is dual to a 2d CFT (or its chiral left-moving sector).

\subsection*{Entropy}

Using the central charge (\ref{cent}) and temperature (\ref{firstlaw}), we find that the Cardy formula for the CFT entropy correctly reproduces the entropy of the cosmological horizon:
\be
S_{CFT} = \frac{\pi^2}{3} T_L c_L = S_{cosm} \ .
\ee
This is the appropriate temperature to use in the Cardy formula because it is the chemical potential conjugate to angular momentum, which is the zero mode of the Virasoro algebra.

As discussed in the introduction, this application of the Cardy formula is speculative.  There is no conclusive evidence that quantum gravity in a de Sitter background is in fact unitary, given that it only appears as a metastable vacuum in string theory \cite{Kachru:2003aw}. Furthermore, a positive central charge does not guarantee the unitarity of a two-dimensional conformal field theory. Finally, it is not understood how the rotating Nariai geometry maps to a thermal state in the CFT.\footnote{We thank A. Strominger for discussions on this point.}  Therefore, although satisfying, the above formula is somewhat numerological and requires further explanation. The agreement persists straightforwardly if we consider adding electric and magnetic charges, as is shown in appendix \ref{charged}.

If the above picture is reasonable, it implies that we should treat the cosmological horizon as any black hole horizon: a thermal state in the dual field theory.

\section{Scalars in Rotating Nariai}\label{scalars}

We now briefly point out the difference in the behavior of scalars in the NHEK and rotating Nariai geometries. In both cases, the scalar field can be separated into a product of spheroidal harmonics and radial functions, $\Psi = R(r)S(\theta)e^{-i\omega t+i m \phi}$. In the rotating Nariai metric (\ref{staticc}), the wave equation for a scalar of mass $\mu$ is \cite{Tachizawa:1992ue}
\be
{1\over \sin\theta}\p_\theta\left(\Delta_\theta \sin\theta \p_\theta S\right) - \left({(m\Omega_c a \sin^2\theta - \Xi m)^2\over \Delta_\theta \sin^2\theta} + a^2 \mu^2 \cos^2\theta -j_{lm}\right)S = 0
\ee
\be
(r^2-1)R'' + 2 r R' + \left({(\omega + k m r)^2\over r^2 - 1} + \tilde{j}_{\ell m}\right)R = 0
\ee
where
\be
\tilde{j}_{\ell m} = {r_c(j_{\ell m} + r_c^2 \mu^2)\over b(a^2 + r_c^2)} \ ,
\ee
and $j_{\ell m}$ is a separation constant the must be determined numerically.  The radial equation can be solved exactly in terms of Whittaker functions, but we need only the large $r$ behavior,
\be
R(r) \sim r^{-\half \pm \beta_{Nariai}} \ , \quad
\beta^2_{Nariai} = {1\over 4} - k^2 m^2 - \tilde{j}_{\ell m} \ . 
\ee
The exponent $\half + \beta$ is the dimension of the dual operator, which is complex for generic mass and angular momentum.

In NHEK, generalizing \cite{Bardeen:1999px} to include a cosmological constant, 
\be
R(r) \sim r^{-\half \pm \beta_{NHEK}} \ , \quad
\beta^2_{NHEK} = {1\over 4} - k^2 m^2 + \tilde{j}_{\ell m} \ .
\ee
Scalars in NHEK are equivalent to charged scalars in an electric field in $AdS_2$.  The NHEK conformal dimension $\half + \beta_{NHEK}$ can be complex for certain values of the parameters and angular momentum, corresponding to the possibility of Schwinger pair production in $AdS_2$ \cite{Pioline:2005pf,Kim:2008xv}.

In rotating Nariai, on the other hand, the weight $\half + \beta_{Nariai}$ is always complex for large enough $\mu^2$, as in the dS/CFT correspondence. The complexity of the conformal weight may indicate that the theory is non-unitary. A possibility for the theory to be rendered unitary would be to consider the unitary principal series representation rather than the highest weight representation as in \cite{Guijosa:2003ze,Lowe:2004nw}. The complexity of the conformal weight is related to cosmological particle production given that $r$ becomes a timelike coordinate for large enough values.

\section*{Acknowledgements}
It is a great pleasure to thank F. Denef and A. Strominger for useful conversations.  This work was supported in part by
DOE grant DE-FG02-91ER40654.

\appendix

\section{Penrose Diagram}
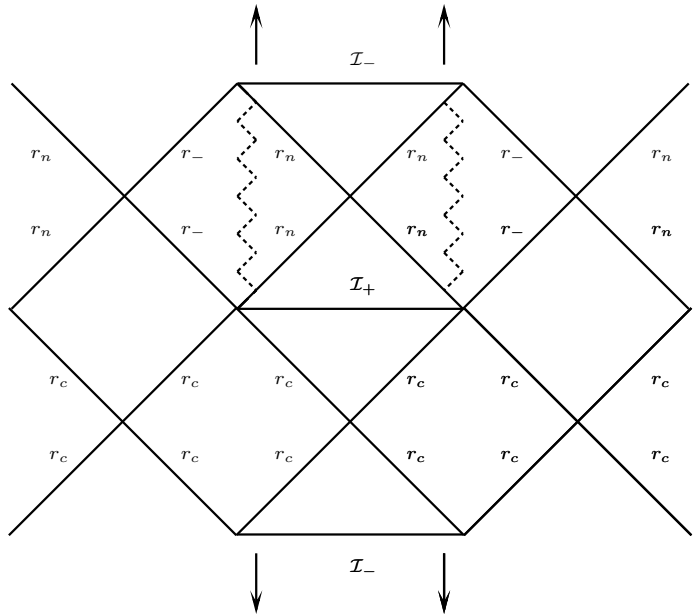
\label{Penrose}

\begin{center}
 \input{NariaiKerrdS.TpX}
 \end{center}
The Penrose diagram of Kerr-de Sitter space for coincident cosmological and black hole horizons. The dotted lines indicate the singularity, $r_n$ is the negative root of the equation $\Delta_{\hat{r}} = 0$, and the left and right sides of the diagram are identified.

\section{Charged Rotating Nariai}\label{charged}

In this appendix we extend the computation of the asymptotic symmetry group and Virasoro central charge to include electromagnetic charge, along the lines of \cite{Hartman:2008pb}. The charged rotating Nariai solution is found as a near horizon limit of the Kerr-Newman-de Sitter black hole with coincident black hole and cosmological horizons, which is a solution to Einstein-Maxwell gravity with a positive cosmological constant
\be
S_{grav} = \frac{1}{16\pi } \int d^4 x \sqrt{-g} \left( R - 2\Lambda  - \frac{1}{4}F^2\right)
\ee
The metric with electric charge $q_e$ and magnetic charge $q_m$ is given by \cite{Booth:1998gf}
\be
ds^2 = \Gamma[\theta]\left[ -(1-r^2)dt^2 + \frac{dr^2}{1-r^2} + \alpha(\theta) d\theta^2 \right] + \gamma(\theta) \left( d\phi + k r dt \right)^2
\ee
where
\be
\Gamma(\theta) = \frac{\rho_c^2 r_c}{b_q(r^2_+ + a^2)}, \quad \alpha(\theta) = \frac{b_q(r^2_c + a^2)}{\Delta_\theta r_c}, \quad \gamma(\theta) = \frac{\Delta_\theta (r_c^2 + a^2)^2\sin^2\theta}{\rho^2_+ \Xi^2}
\ee
and $\phi$ is an angular coordinate with periodicity $\phi \sim \phi + 2\pi$. We have further defined
\be
\rho_c^2 = r_c^2 + a^2 \cos^2\theta, \quad b_q = \frac{6r_c^2/\ell^2 + 3 r_c^4/\ell^4+q^2/\ell^2-1}{r_c(r^2_c + a^2)(1+r_c^2/\ell^2)}, \quad k = -\frac{2 a r_c^2\Xi }{b_q(r_c^2 + a^2)^2} \ ,
\ee
with $q^2 = q_e^2 + q^2_m$. The gauge field is given by
\be
A = f(\theta)\left( d\phi + k r dt \right)
\ee
with
\be
f(\theta) = \frac{(r^2_c + a^2)[q_e(r^2_c - a^2\cos^2\theta)+2 q_m a r_c \cos\theta]}{2\rho_c^2 \Xi a r_c} \ .
\ee
It is a straightforward extension of the non-charged case to find that in the limit where the black hole and cosmological horizons coincide, the first law for the cosmological horizon reads
\be
d S_{cosm} = \beta_L d \mathcal{Q}_{\partial_\phi} + \Phi_e d\mathcal{Q}_e + \Phi_m d\mathcal{Q}_m
\ee
where $\beta_L = 2\pi |k|$. The computation of the asymptotic symmetry group proceeds as before, resulting in a single Virasoro algebra with central charge
\be\label{appac}
c_L = {12 r_c \sqrt{(r_c^2-3r_c^4/\ell^2 - q^2)(1+r_c^2/\ell^2)}\over -1 + 6 r_c^2/\ell^2 + 3 r_c^4/\ell^4-q^2/\ell^2} \ ,
\ee
and a similar entropy matching formula
\be
S_{CFT} = \frac{\pi^2}{3}c_L T_L = S_{cosm} \ .
\ee

\end{document}